\documentclass[11pt]{article}

\usepackage{amssymb,amsmath, epsfig}
\usepackage{changepage}
\usepackage{caption}
\usepackage[section]{placeins}

\def\u{\vskip  .075 in}
\def\nh{\noindent\hangindent=1 true cm \hangafter = 1}

\def\nh{\noindent\hangindent=1 true cm \hangafter = 1}

\def\u{\vskip  .1 in}

\def\B {\begin{eqnarray*}}

\newcommand{\bel}[1]{\begin{equation}\label{#1}}

\newcommand{\be}{\begin{equation}}
\newcommand{\qe}{\end{equation}}
\newcommand{\ee}{\end{equation}}
\newcommand{\baS}{\begin{eqnarray}}
\newcommand{\ba}{\begin{eqnarray}}
\newcommand{\ea}{\end{eqnarray}}

\def\Na{$g_{Na,max}$}

\def\IKDR{$g_{KDR,max}$}

\def\EN{\end{eqnarray*}}

\begin{document}
\title{Simplified models of pacemaker spiking in raphe and locus coeruleus neurons\\
\   \\
{\normalsize
Henry C. Tuckwell$^{1\dagger, *}$,  Ying Zhou$^{2\dagger}$ , Nicholas  J. Penington $^{3,4\dagger}$\\   \
\  \\ 
 \              \\
$^1$ School of Electrical and Electronic Engineering, University of Adelaide,\\
Adelaide, South Australia 5005, Australia \\
 \              \\
$^2$ Mathematical Biosciences Institute, Ohio State University, 1735 Neil Ave. Columbus, Ohio 43210, USA \\
\              \\
$^3$ Department of Physiology and Pharmacology,\\
$^4$ Program in Neural and Behavioral Science and Robert F. Furchgott
Center for Neural and Behavioral Science \\
State University of New York,
Downstate Medical Center,\\
Box 29, 450 Clarkson Avenue, Brooklyn, NY 11203-2098, USA\\
\     \\
$^{\dagger}$ {\it Emails:} henry.tuckwell@adelaide.edu.au; zhou.494@osu.edu; Nicholas.Penington@downstate.edu\\ 
\     \\
 }}

\maketitle

\newpage 
\begin{abstract}  
Many central neurons, and in particular certain brainstem aminergic neurons
exhibit spontaneous  and fairly regular spiking with frequencies of order 
a few Hz. A large number of ion channel types contribute to such spiking so that
accurate modeling of spike generation leads to the requirement of solving very large
systems of differential equations, ordinary in the first instance. Since analysis
of spiking behavior when many synaptic inputs are active adds further to the
number of components, it is useful to have simplified mathematical models 
of spiking in such neurons so that, for example, stochastic features of
inputs and output spike trains can be incorporated. In this article we investigate
two simple two-component models which mimic features of spiking in
serotonergic neurons of the dorsal raphe nucleus and noradrenergic neurons of the locus coeruleus. The first model is of the Fitzhugh-Nagumo type and the second is a reduced Hodgkin-Huxley model.
For each model solutions are computed with two representative sets
of parameters. Frequency versus input currents are found and reveal
Hodgkin type 2 behavior.  For the first model a bifurcation 
and phase plane analysis supports these findings. The spike trajectories in the second model are very similar to those in DRN SE pacemaker activity but there are more parameters than in the Fitzhugh-Nagumo type model. The article concludes with a brief review of previous modeling of these types of neurons and its relevance to
studies of serotonergic involvement in spatial working memory 
and obsessive-compulsive disorder.
\end{abstract}


\rule{80mm}{.5pt}
\tableofcontents
\rule{80mm}{.5pt}

\noindent {\bf Abbreviations}  \\
\noindent 5-HT, 5-hydroxytryptamine (serotonin); AC, adenylate cyclase; 
AHP, afterhyperpolarization; cAMP, cyclic adenosine monophosphate;
CREB, cAMP response element binding protein; CRN, caudal raphe
nucleus; 
 DRN, dorsal raphe nucleus;  EPSP, excitatory post-synaptic potential; 
 GABA, gamma-aminobutyric acid;  Hz, hertz;  ISI, interspike interval; LC, locus coeruleus; NA, noradrenaline or noradrenergic; 
REM, rapid eye movement; 
 SE, serotonin or serotonergic.


%

  \section{Introduction}
Neurons which exhibit (approximately) periodic spiking in the presumed absence of synaptic input are called  autonomous pacemakers, and include neurons found 
in the subthalamic nucleus, nucleus basalis, globus pallidus, raphe nuclei,
cerebellum, locus ceruleus, ventral tegmental area, and substantia nigra (Ramirez et al., 2011). Some pacemaking cells may require small amounts of depolarizing
inputs, natural or laboratory. Thus, for example Pan et al. (1994) found that all of 42
rat LC neurons fired spontaneously, whereas in Williams et al. (1982) and 
Ishimatsu et al. (1996) it was reported that most cells did not require excitatory input to
fire regularly.  These latter three sets of results were obtained in vitro.
Sanchez-Padilla et al. (2014) reported that spike rate in mouse LC neurons  
was not affected by blockers of glutamatergic or GABAergic synaptic input,
supporting the idea that these cells were autonomous pacemakers. 

In this article we are concerned with certain cells in brainstem nuclei, particularly the DRN and the LC.  In most common experimentally employed animals except cat, the locus coeruleus is almost completely
homogeneous, consisting of noradrenergic neurons which in rat mumber about
1500 (Swanson, 1976; Berridge and Waterhouse, 2003). 
 The number of neurons in the rat DRN is between about 12000 and 15000
(Jacobs and Azmitia, 1992; Vertes and Crane, 1997) of which up to 
50\% are principal serotonergic cells (Vasudeva and Waterhouse, 2014)  but there are also present ~1000 dopaminergic cells (Lowry et al., 2008) and 
 GABAergic cells, whose density varies throughout the divisions of the nucleus,  as well as several other types of neuron. 
 
 In the present article we are mainly concerned with the principal neurons of the
 DRN (serotonergic cells) and LC (noradrenergic cells), which often exhibit a slow regular pattern
of firing with frequencies of order 1 to 2 Hz in slice and sometimes higher
in vivo. Included are midbrain or pons (midbrain)  noradrenergic neurons 
in locus coeruleus and serotonergic neurons of the dorsal and other raphe
nuclei.

  The origins of pacemaker firing differ amongst 
various neuronal types.
Thus, brainstem dopaminergic neurons may  
fire regularly without  excitatory synaptic input (Grace and Bunney
1983; Harris et al., 1989). Underlying the rhythmic activity are subthreshold oscillations
which were demonstrated with a mathematical model to reflect an interplay between an L-type calcium current and a calcium-activated
potassium current (Amini et al, 1999). 
The mechanisms of pacemaker firing in LC neurons are not
fully understood, although there is the possibility it is sustained by
a TTX-insensitive persistent sodium current (De Carvalho et al., 2000;
Alvarez et al., 2002). For serotonergic neurons of the DRN, there have
been no reports of a persistent sodium current and L-type calcium
currents are relatively small or absent (Penington et al., 1991) so the main candidate
for depolarization underlying pacemaking is a combination
of T-type calcium current and  the classical
fast TTX-sensitive sodium current which dominates the pre-spike
interval (Tuckwell and Penington, 2014).  In some cells the hyperpolarization-activated cation current may also play a role. 

In Figure 1 are shown portions of spike trains in mouse and rat LC and
rat DRN and CRN. It is noteworthy that the frequency of firing of LC and DRN 
principal neurons depend on sleep stage. Thus for example,  in rats, waking ,
slow-wave sleep and REM sleep are accompanied by LC firing rates
of about 2.2 Hz, 0.7 Hz and 0.02 Hz, respectively (Foote et al., 1980;
Aston-Jones and Bloom, 1981; Luppi et al., 2012). 
Figure 2 shows  computed spikes for the detailed model of rat DRN SE neurons
 of Tuckwell and 
Penington (2014) with 4 different parameter sets. In this model the main
variables are membrane potential and intracellular calcium ion concentration
which satisfy ordinary differential equations. There are, however, 
11 membrane currents which drive the system which results in a system with
18 components and over 120 parameters. In order to study quantities like 
interspike interval distributions with various sources of random synaptic input,
it is helpful to have a simpler system of differential equations which might
yield insight into the properties of the complex model whose execution
with random inputs over hundreds of trials would be overly time consuming.
It is also pointed out that LC and DRN principal neurons are each
responsive to activation of about 20 different receptor types,
making computational tasks even more cumbersome with an 18-component model
(Kubista and Boehm, 2006; Maejima et al., 2013). 
A simplified model of spike generation would be useful in modeling the
dynamics of serotonin release and uptake (Flower and Wong-Lin, 2014).

     \begin{figure}[!h]
\begin{center}
\centerline\leavevmode\epsfig{file=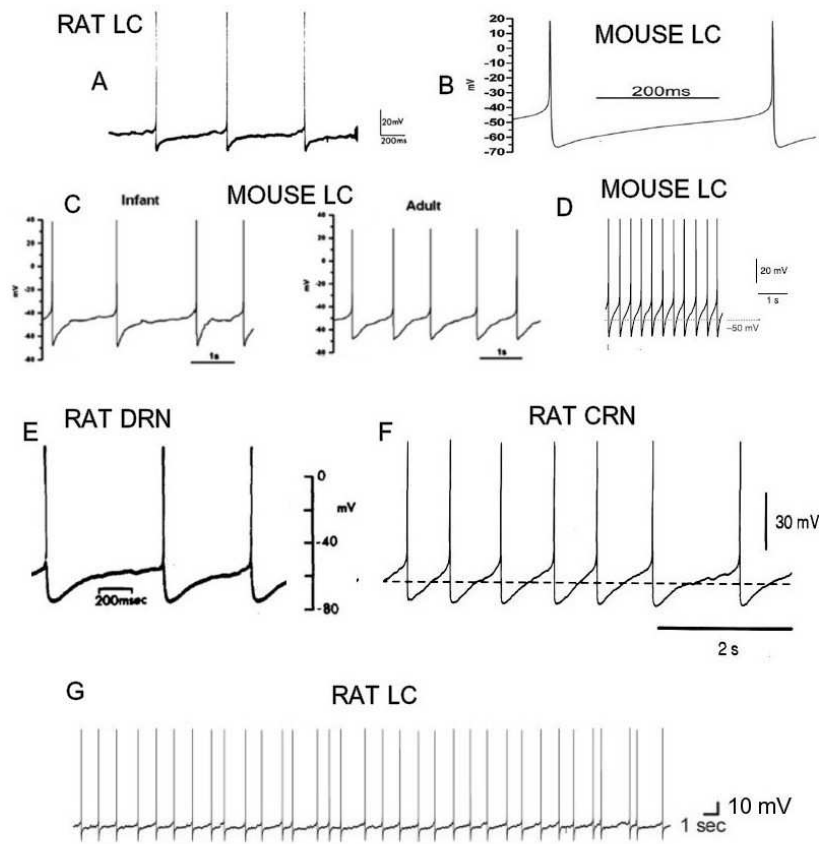,width=4.25in}
\end{center}
\caption{Some representative spikes from rat and mouse raphe nuclei and LC neurons.
{\bf A}. Part of a train of spikes in a rat LC neuron in slice. Markers 20 mV and 200 ms. (Andrade and Aghajanian, 1984). {\bf B}.  Detail of the course of the average membrane potential
in a mouse LC neuron during an interspike interval. (De Oliveira et al., 2010). {\bf C}. Action potentials in infant  (7 to 12 days) and adult (8 to 12 weeks) mice.
(De Oliveira et al., 2011). {\bf D}. Whole-cell current-clamp recording of spikes in a
 mouse (21 to 32 days) LC neuron. (Sanchez-Padilla et al., 2014). {\bf E}.
A few spikes from rat dorsal raphe nucleus (slice), (Vandermaelen and Aghajanian, 1983). {\bf F}. Portion of a spike train from rat caudal raphe nucleus. (Bayliss et al., 1997).  {\bf G}. Train of spontaneous spikes at a mean frequency of 0.85 Hz for a rat LC
neuron in vitro. (Jedema and Grace, 2004).} 
\label{fig:wedge}
\end{figure}

      \begin{figure}[!t]
\begin{center}
\centerline\leavevmode\epsfig{file=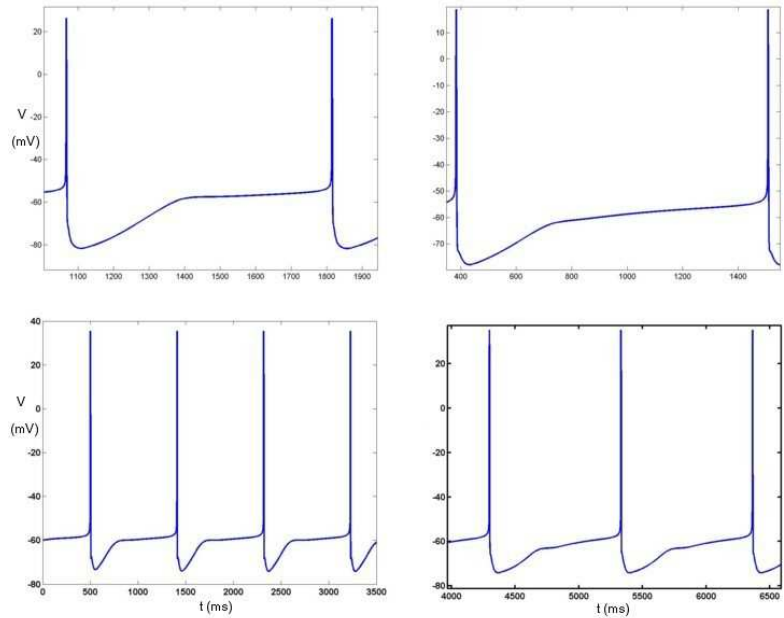,width=4.5in}
\end{center}
\caption{Examples of computed spikes in a model for serotonergic neurons of the rat
dorsal raphe nucleus from the model of Tuckwell and Penington (2014).
Illustrated are the prolonged afterhyperpolarizations after a spike. The
subsequent climb to threshold is plateau-like, sometimes being almost
horizontal. Membrane potentials in mV are plotted against time in ms.} 
\label{fig:wedge}
\end{figure}

\section{Description of a two-component model}
A first goal was to construct a two-component differential equation model 
whose solutions for the membrane potential
 broadly mimicked those for the muti-dimensional model developed 
in Tuckwell and Penington (2014), which gave satisfactory agreement
with experimental voltage  (see references in the preceding reference).

To this end the following pair of equations was found, with suitable choice
of parameters, to have soutions with the desired properties. Here $V$, in mV,  is the
depolarization of membrane potential from resting value and $R$ is a 
recovery variable. In keeping with the properties of the variables in the
Fitzhugh-Nagumo equations (see for example Tuckwell, 1988, Section 8.8)
$R$ will be (arbitrarily) ascribed units of mV/ms.
Then we have
\be \frac{dV}{dt} =  \frac{1}{\alpha}(V-V_1)(V-V_2)(V_3-V) - \lambda R + I_{App}, \ee
\be  \frac{dR}{dt} = \frac{\epsilon} { 1 + \exp\big[\frac{-(V - V_a)}{k_a} \big] } 
     + kRV, \ee
which system is usually to be solved with the initial conditions 
$V(0)=V_0, R(0)=R_0$.  The initial value of $V$ is usually set at the
resting membrane potential so that  $V_0=V_R$ whose average value for the cells
of interest is -64.4 mV (Tuckwell, 2013).

For the application to spiking, it is assumed that the parameters $\alpha$,
$\lambda$, $\epsilon$, $k$ and $k_a$ are all positive. 
 The zeros of the cubic
\be f(V)=  \frac{1}{\alpha} (V-V_1)(V-V_2)(V_3-V) \ee
are chosen such that $V_1<V_2<V_3$, with $V_1<0$ and $V_2<0$.

\section{Examples of results for the two-component model}
In this section we give examples of computed solutions for the above two-component system
and consider some of the properties of solutions. 

\subsubsection*{Examples with two sets of parameters} 
There are ten parameters for the system of equations (1) and (2). 
Here we describe solutions for two sets of parameters, whose values are
listed in Table 1. The solutions for both sets consist of periodic solutions in
$V$ and $R$ where the first component mimics trains of action potentials.

For parameter set 1 the (numerical) solutions are depicted in Figure 3 and 
some of the details of the solutions, such as maximum and minimum values of
$V$, mean ISI, mean duration measured at $V=-40mV$ and maximum value of $R$  are given in the second column of Table 2. These results were obtained using
an Euler scheme with $\Delta t=0.02$ ms.

 The voltage trajectories resemble in form
some of the experimental ones in Figure 1, particularly for rat and mouse LC neurons (A and C), and to a lesser extent the rat DRN neurons (E). The duration of
the action  potential (measured at -40 mV) is only 0.55 ms, which is
too short for these brainstem neurons.  Furthermore, the AHP declines to
a very low value of about -109 mV which is 49 mV below the assumed resting level.
In order to make the minimum considerably higher in accordance with most 
experimental values, it is noted that the minimum of $V$ occurs when
$dV/dt=0$ or when \be f(V)=\lambda R^* -I_{App}\ee where $R^*$ is the
value of $R$ when the minimum of $V$ occurs. The graphical situation
is shown in Figure 4. By judicious choice of values of the parameters, it
was possible to obtain periodic solutions with minima of $V$ at an 
appropriate value of about -83 mV. This resulted in the parameter set 2
whose solution properties are displayed in column 3  of Table 2. These results were
also obtained using an Euler scheme with step 0.02. The corresponding
results with a step of $\Delta t = 0.005$ ms are given in column 4. 
 The latter much smaller value of $\Delta t$ should lead to
more accurate solutions, but considering that the computing time was
about 30 times longer and the change in solution properties such as
ISI and duration were less than 1\%, it is satisactory to use the larger time step. 
A comparison was also made with results obtained using a 4th order Runge-Kutta 
which was an order of magnitude slower than the Euler method and gave an
almost identical ISI of 869.04 ms.  

\begin{center}

\begin{table}[h]
    \caption{Two sets of parameters for the system (1),(2)}
\smallskip
\begin{center}
\begin{tabular}{llllll}
  \hline
     Parameter    &  Set 1 & Set 2 & Parameter & Set 1 & Set 2\\
  \hline
$\alpha$ & 400 & 400 & $V_1$ & -77.4  & -60\\
$\epsilon$  & 30 & 5  & $V_2$ & -61  & -50  \\
$k_a$  & 2 &2   & $V_3$ & 20 & 20 \\
$V_a$ & -10  & -10  & $I_{App}$ & 15 & 15 \\
$\lambda$ & 60 & 20 &  $k$ & 0.00042   &       0.0000525\\

      \hline
\end{tabular}

\end{center}

\end{table}

\end{center}

      \begin{figure}[!h]
\begin{center}
\centerline\leavevmode\epsfig{file=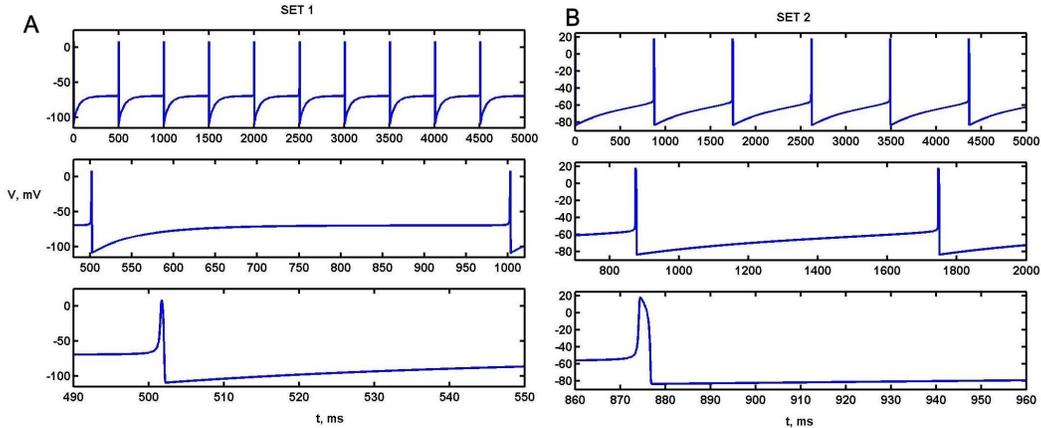,width=5.5in}
\end{center}
\caption{A. Plots of numerical solutions for the two-component model with the
first parameter set of Table 1. $V(t)$ is plotted against $t$ with
three different timescales to show details of spike train and spike.
B. Corresponding results for set 2 parameters.} 
\label{fig:wedge}
\end{figure}

%
%
%

\begin{center}

\begin{table}[h]
    \caption{Some details for the spike trains for set 1 and set 2 parameters with $I_{App}=15$.}
\smallskip
\begin{center}
\begin{tabular}{lccc}
  \hline
     Property   &  Set 1  ($\Delta t=0.02$  ms) & Set 2 ($\Delta t=0.02$ ms) &
Set 2 $\Delta t=0.005$ ms  \\ 
  \hline
Mean ISI (ms) & 501.1 & 870.8 & 869.5  \\
Mean Duration (ms) & 0.55  & 2.81 & 2.79 \\
Max (V) (mV) & +8.9  & +18.7 & +18.5\\
Min(V) (mV) & -109.4 mV & -83.5 & -83.4 \\
Max(R) & 8.7 & 10.96 & 10.90 \\

      \hline
\end{tabular}

\end{center}

\end{table}

\end{center}
.
\subsection{Frequency versus current curves}
The above results for sets 1 and 2 parameters were all obtained with the
$I_{App}=15$. It is of interest to compute frequency of action potentials
for various applied currents, as this corresponds to certain experimental data.
The plot of output frequency versus applied current is called an f/I curve which
differs in its characteristics from neuron to neuron. Often there is a threshold
current for action potentials, which in the classical literature was called the rheobase
current.
      \begin{figure}[!h]
\begin{center}
\centerline\leavevmode\epsfig{file=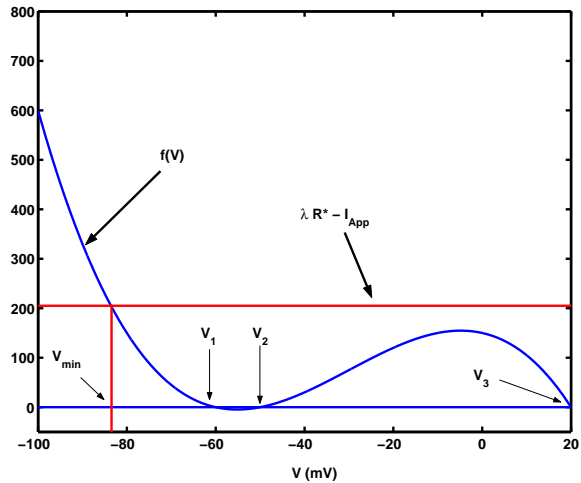,width=3in}
\end{center}
\caption{Illustrating the graphical solution for the minimum of $V$.} 
\label{fig:wedge}
\end{figure}

Hodgkin (1948)
found for squid axon preparations that there were two broad types of f/I curves. Type 1 consisted of
an f/I curve that smoothly rose from zero at a particular value of the
input current, whereas type 2 curves discontinuously rose at a 
certain threshold current. Tateno et al. (2004) found that regular spiking and fast spiking neurons in the rat
somatosensory cortex exhibit Type 1 and Type 2 firing behaviours, respectively.  Mathematical explanations for the two types of threshold are found in the nature of the bifurcation which
accompanies the transition from rest state to a periodic firing mode as discussed in Section 4.

Graphs of the frequency of repetitive spiking versus depolarizing input current
are shown in Figure 5 for parameter sets 1 and 2. In both cases,  at a particular value of the applied current $I_{App}$ the
frequency jumps from zero to a positive value. 
For parameter set 1, the threshold current for firing is very close to 15 at which the firing frequency is about 2 Hz. For parameter set 2, the threshold curent is about
$I_{App}=4.7$ at which the friring frequency jumps from zero to about
0.29 Hz. Thus the responses of the model with either parameter set
are those of type 2 neurons (Hodgkin, 1948).
The nature of the f/I curves for the approximate model is thus similar to that for  both experimental results for 
DRN SE neurons and  for the multi-component model (Tuckwell and Penington, 2014). 
%

\subsubsection{Autonomous pacemaker activity}
The above results on frequency versus current have indicated that
to make the model neuron fire with parameter sets 1 and 2, current $I_{App}>0$ must be applied. If $R(0)=0$ this is  necessary because the cubic $f(V)$ defined in 
Eq. (3) is negative for a range of values of $V$, and in particular
the range containing $V(0)$. If $V_1=V_2=-60$, the cubic is tangential to the
$V$-axis, and $f(V)$ is never negative. Then firing, albeit very slow, was demonstrated to occur
for values of $I_{App}$ extremely close to zero, which would imply
autonomous firing in the limit. It is possible by choosing a cubic $f(V)$ with
only one real root, for example at $V_3=20$ (as in parameter sets 1  and 2), 
so that $f(V)>0$ for all  $V<V_3$, in which case autonomous firing
could occur; that is with $I_{App}=0$.  The resulting source function would then be similar to the
steady state curve in Figure 18 of the multidimensional model (Tuckwell and
Penington, 2014) where in some cases it was found that pacemaker firing
occurred in some cases for $I_{App}=0$, or even $I_{App}<0$ whereas
in others a small depolarizing drive was necessary.

      \begin{figure}[!t]
\begin{center}
\centerline\leavevmode\epsfig{file=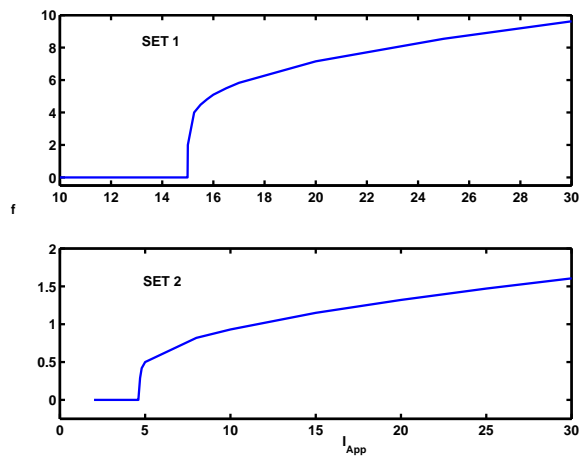,width=3in}
\end{center}
\caption{Frequency versus input depolarizing current for parameter sets 1 (above)
and 2 (below).} 
\label{fig:wedge}
\end{figure}

\section{Effects of changing parameters and phase plane analysis}
A detailed compilation of the effects on the spike train properties as
each of the ten parameters is varied is not explored here. Instead, 
Table 3 lists the effects of increasing or decreasing each of the ten parameter values relative to their values in set 2 on the mean ISI, mean duration of spike, maximum and minimum of voltage during spike, and maximum value of the recovery variable during spiking. We focus attention on the ISI and spike duration. 

For the ISI,  significantly altering each of the parameters $\epsilon$, $\lambda$, $V_a$ and $k_a$
had little or no effect.  On the other hand, the ISI was increased by
decreasing any of $\alpha$, $I_{App}$, $V_1$ and $k$ or by increasing
either $V_2$ or $V_3$.  The duration of spikes was little affected
by significant changes in any of $k$, $k_a$, $I_{app}$, $V_1$,
$V_2$ or by decreases in $V_a$, although increases in $V_a$
led to a substantial increase in duration.  Significant increases in duration resulted from decreases in any of $\alpha$, $\epsilon$ or $\lambda$, or increases in $V_3$.

\begin{center}
\begin{table}[h]
    \caption{Properties of the spike trains for different sets of parameters relative to set 2}
\smallskip
\begin{center}
\begin{tabular}{lccccc}
  \hline
     Parameters   &  Mean ISI & Mean Duration & Max(V)
 & Min(V) & Max(R) \\ 
  \hline
Set 2 & 869.04  & 2.74 & 18.37 & -83.40 & 10.88\\
$\alpha = 2000$ & 462.4  & 3.0822 & 0.26 & -91.92& 4.53\\
$\alpha = 200$ & 1231.84 & 4.0267 & 19.69 & -81.73 & 18.32\\
$\epsilon = 2$ & 849.32  & 5.47 & 19.84 & -82.15 & 9.87 \\
$\epsilon = 8$ & 884.04 & 2.005 & 17.01 & -84.32 & 11.66 \\
$\lambda = 10$ & 853.02 & 4.58 & 19.58 & -82.40 & 20.14 \\
$\lambda = 30$ & 881.76 & 2.085 & 17.23 & -84.18 & 7.70\\
$I_{app} = 10$ & 1069 & 2.74 & 17.95 & -83.40 & 10.63 \\
$I_{app} = 20$ & 755.52 & 2.74 & 18.78 & -83.40 & 11.13 \\
$V_1= -65$ & 1127.82 & 2.8667 & 18.59 & -86.82& 11.51\\
$V_1= -55$ & 794.7 & 2.66 & 18.10 & -80.15 & 10.27\\
$V_2 = -55 $ & 771.76 & 2.812 & 18.62 & -86.21 & 11.65 \\
$V_2 = -45 $ & 1128.26 & 2.7067 & 18.05 & -80.78 & 10.14 \\
$V_3 = 15$ & 815.24 & 2.52 & 13.10 & -81.73 & 9.12 \\
$V_3 = 25$ & 919.14 & 3.025 & 23.63 & -84.99 & 12.80\\
$V_a = -20$ & 883.14 & 2.63 & 17.78 & -84.23 & 11.59\\
$V_a = 0$ & 840.84 & 3.14 & 18.86 & -81.82 & 9.62\\
$k_a = 1$ & 869.3 & 2.73 & 18.37 & -83.42 & 10.90\\
$k_a = 3$ & 868.76 & 2.75 & 18.36 & -83.38 & 10.87\\
$k = 0.0000325$ & 1396.54 & 2.74 &18.37 & -83.42 & 10.89\\ 
$k = 0.0000725$ & 632.26 & 2.74 & 18.37 & -83.39 & 10.88\\
      \hline
\end{tabular}
\end{center}
\end{table}
\end{center}
\newpage
 To see the bifurcations involved when the injected current $I_{App}$ is changed, we can analyze the phase plane of the model.
The two nullclines of the two-component model are 
\be  \label{firstNullcline} R = \frac{1}{\lambda \alpha}(V-V_1)(V-V_2)(V_3-V) + \frac{I_{App}}{\lambda}, \ee
for $dV/dt = 0$, 
and 
\be  R = -\frac{\epsilon} { 1 + \exp\big[\frac{-(V - V_a)}{k_a} \big] } 
     \cdot \frac{1}{kV}, \ee
for $dR/dt = 0$.
\begin{figure}[!ht].
\begin{center}
\centerline\leavevmode\epsfig{file=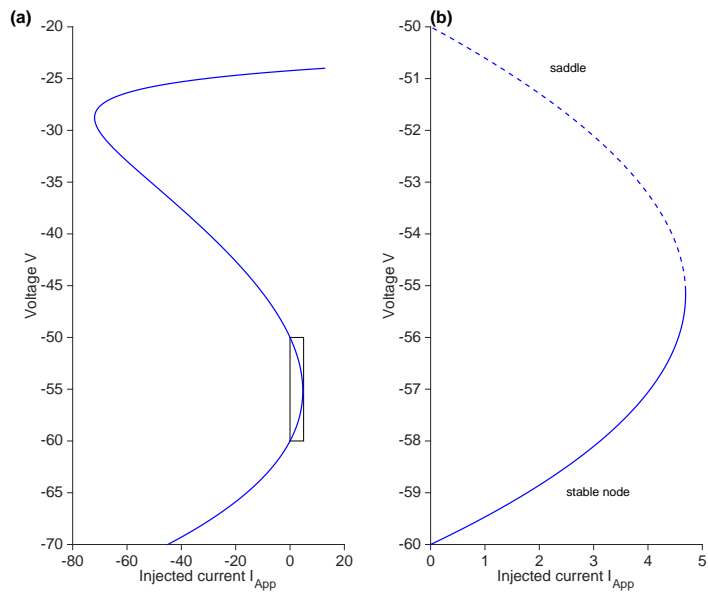,width=4.25in}
\end{center}
\caption{Saddle-node bifurcation diagram of the two-component model. In panel (a), voltage $V$ of the equilibria are plotted with respect to the injected current $I$. A close-up picture of the rectangular area in panel (a) is plotted in panel (b), where the dashed branch represents the voltage-current relationship for the saddle, and the solid branch represents that of the stable node. The parameters other than the injected current $I_{App}$ are the same as set 2 in Table 1, except $\epsilon = 8$.} 
\label{fig:saddleNodeBifurcation}
\end{figure}
Depending on the parameters, the system may have one equilibrium or three equilibria. The voltage variable, $V$, for each equilibrium, satisfies the equation
\begin{equation}\label{equilibriumEqn}
\frac{1}{\lambda \alpha}(V-V_1)(V-V_2)(V_3-V) + \frac{I_{App}}{\lambda} = -\frac{\epsilon} { 1 + \exp\big[\frac{-(V - V_a)}{k_a}\big] }. 
\end{equation}
Solving equation (\ref{equilibriumEqn}) for the applied current $I_{App}$, we find that the system goes through a saddle-node bifurcation when $I_{App}$ increases. In Figure (\ref{fig:saddleNodeBifurcation}a), the voltage $V$ in the equilibrium equation (\ref{equilibriumEqn}) is plotted with respect to the applied current $I_{App}$. As we can see from Figure (\ref{fig:saddleNodeBifurcation}a), when $I_{App} = 0$, $\epsilon = 8$, and the other parameters are the same with parameter set 2, the system has three equilibria, visualized in Figure (\ref{fig:phasePortraits}a) by the intersection of the cubic $V$-nullcline and the $R$-nullcline. In the lower voltage range, there is a stable node accompanied by a nearby saddle. The stable node corresponds to the resting state, and the saddle sets a threshold for the initial voltage required for there to be a (non-repetitive) spike. As the current $I_{App}$ increases, the distance between the stable node and the saddle decreases. Eventually the saddle and the node collide and annihilate each other through a saddle-node bifurcation, and the bifurcation diagram is as in  Figure (\ref{fig:saddleNodeBifurcation}b). Once $I_{App}$ is larger than the bifurcation value, there is only one equilibrium which is an unstable focus, and the system has a limit cycle (see Figure \ref{fig:phasePortraits}b where $I_{App} = 15$).

\begin{figure}[!hb]
\begin{center}
\centerline\leavevmode\epsfig{file=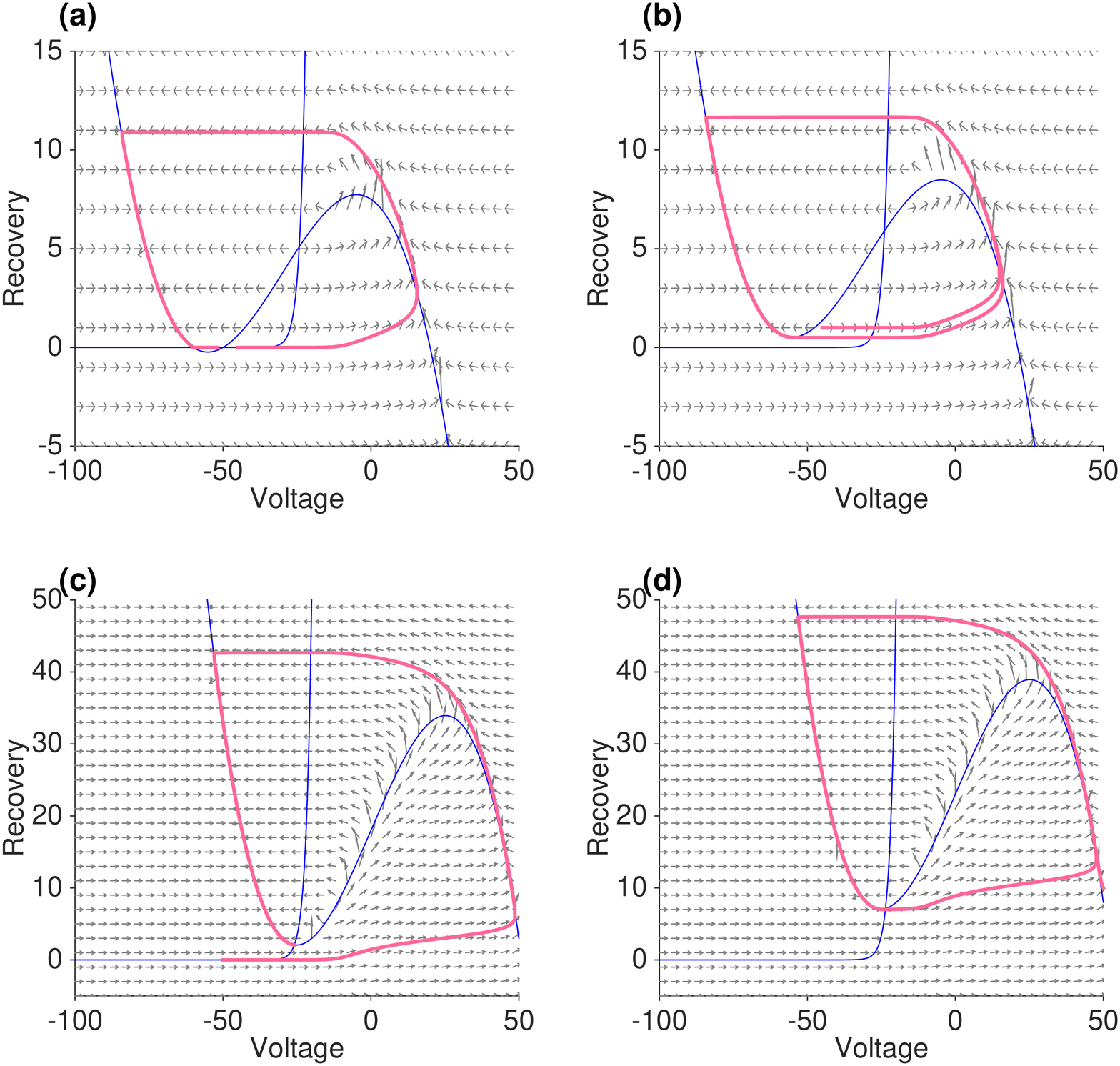,width=5.25in}
\end{center}
\caption{Phase portraits of the two-component model for different parameters. In panel (a), $I_{App} = 0$, $\epsilon = 8$, while the other parameters are the same as set 2 in Table 1, and the resting state is a stable node. As $I_{App}$ increases, the system goes through a saddle-node bifurcation. In panel (b), $I_{App}$ is increased to $15$, the system has only one equilibrium remaining, and there is a limit cycle. In panel (c), $V_1 = -30$, $V_2 = -20$, $V_3 = 50$, and other parameters coincide with those in panel (b). With this parameter set, the system has only one equilibrium, which is a stable node. As $I_{App}$ is increased, the system goes through an Andronov-Hopf bifurcation, and the stable node loses stability to give birth to a limit cycle, such as the one in panel (d). In panel (d), $I_{App} = 40$ while the other parameters are the same as in panel (c).} 
\label{fig:phasePortraits}
\end{figure}

The system may also go through an Andronov-Hopf bifurcation when $I_{App}$ increases. For example, when $I = 15$, $\epsilon = 8$, $V_1 = -30$, $V_2 = -20$, $V_3 = 50$, and all other parameters coincide with parameter set 2, then the system has only one equilibrium, which is a stable focus corresponding to the resting state (see Figure \ref{fig:phasePortraits}c). When $I_{App}$ is increased, the stable focus loses its stability, a limit cycle comes to exist (see Figure \ref{fig:phasePortraits}d), and the model exhibits spiking behavior.

\par

From the theory of dynamical sytems (Ishikevich, 2007)
spiking as a sequitur to either a saddle-node bifurcation or an Andronov-Hopf bifurcation results in a neuron with Type 2 dynamics as was found in the numerically generated f/I curves of Figure 6.

\section{A reduced physiological model with two component currents}
With the multi-component neuronal model for DRN SE neurons (Tuckwell and Penington, 2014) there are 11 component currents and a variety of solution behaviors for different choices of
parameters. It is often difficult to
see which parameters are responsible for various spiking properties. 
Some parameter sets lead to satisfactory spike trains with 
properties similar to experiment, but sometimes spike durations are
unacceptably long.  Other sets give rise to 
spontaneous activity but with spikes which have uncharacteristic large notches on
the repolarization phase. Doublets and triplets are also often 
observed, and although these are sometimes found in experimental spike trains,
it is desirable to find solutions depicting the regular singlet spiking and
relatively smooth voltage trajectories usually observed.

In the previous section we have described a simplified two-component
model whose first component representing membrane potential
did resemble pacemaker spiking in certain brainstem neurons.
An alternative model, with more parameters, can be constructed
using full descriptions for the fast sodium transient currrent
 $I_{Na}$ and the
delayed rectifier potassium current $I_{KDR}$ as originally in the Hodgkin-Huxley 
(1952)  model, but not including the leak current which may be
approximately absorbed
into an applied current $\mu$. 
The emphasis is again on the membrane potential trajectories.

The basic differential equation for the voltage is 
\be  C\frac{dV}{dt}=-[I_{KDR}  + I_{Na} + \mu] \ee 
where C is capacitance,  and $\mu$ is an added current in nA which 
depolarizes if negative.  The initial value of $V$ is again taken to be
resting potential, $V_R$. 
All potentials are in mV, all times are in ms, all conductances are in $\mu$S  and  capacitance is in nF.

Activation and inactivation variables are determined by the 
first order equations
\be \frac{dm}{dt} = \frac{m_{\infty} - m}{\tau_m} \ee
\be \frac{dh}{dt}=\frac{h_{\infty} - h}{\tau_h} \ee
where $m_{\infty}$ and $h_{\infty}$ are steady state values which depend on voltage.
The quantities $\tau_m$ and $\tau_h$ are time constants which may also
depend on  voltage.

The fast transient sodium current,  $I_{Na}$ is given by the classical form 
\be I_{Na}=g_{Na,max} m_{Na}^3h_{Na}(V-V_{Na}) \ee
with activation variable $m_{Na}$ and inactivation $h_{Na}$. 
For the steady state activation we put 
\be m_{Na, \infty} = \frac{1}{1 + e^{-(V - V_{Na_1})/k_{Na_1} } } \ee
with corresponding time constant 
\be \tau_{m, Na}=    a_{Na} + b_{Na}e^{- \big( (V - V_{Na_2})/k_{Na_2}\big)^2}, \ee
which fits well the forms used by some authors
(McCormick and Huguenard, 1992; Traub et al., 2003)  but not all. 
The steady state inactivation may be written 
\be h_{Na, \infty}= \frac{1}{1 + e^{(V -V_{Na_3})/k_{Na_3} } } \ee
with corresponding time constant  fitted with
\be \tau_{h, Na}=   c_{Na} + d_{Na}e^{- \big( (V - V_{Na_4})/k_{Na_4}\big)^2}. \ee

We take the  form for the delayed rectifier potassium
 current to be   
\be I_{KDR}= g_{KDR,max} n^{n_k}(V-V_{KDR}) \ee 
where $n$ (the traditional symbol)  is the activation variable which satisfies a differential
equation like (6) and $V_{KDR}$ is the reversal potential.
The steady state activation is written 
 \be n_{\infty}=\frac{1} { 1 + e^{- (V - V_{KDR_1})/k_{KDR_1}} }  \ee
and the time constant 
 \be \tau_n= a_{KDR} +  \frac {b_{KDR}} { \cosh( (V - V_{KDR_2})/k_{KDR_2}   )}   \ee

\subsection{Examples of pacemaker-like firing with two parameter sets}

We illustrate the solutions with two choices of parameter sets.  Set 1 is adopted from the parameters used for fast sodium and delayed rectifier potassium in the 11-current component model  (Tuckwell and Penington, 2014) and the other is based in largely on values in Belluzi and Sacchi (1991) which we denote  by Set 2. 
For the latter we also use the resting potential 
and the cell capacitance from Kirby et al. (2003), both being for rat DRN SE cells. The two sets of parameters
are summarized in Table 4.

\begin{center}
\begin{table}[!ht]
    \caption{Two basic parameter sets for the Na-K system}
\smallskip
\begin{center}
\begin{tabular}{lcc}
  \hline
 Parameter &  Set 1 &  Set 2 \\
\hline 
    $V_{Na_1}$ &   -33.1 &   -36\\
 $k_{Na_1}$ & 8 & 7.2\\
 $V_{Na_3}$ & -50.3 & -53.2 \\
 $k_{Na_3}$ & 6.5 & 6.5 \\
$V_R$ & -60 & -67.8 \\
$\tau_{m,Na_c}$  & 0.2 & 0.1 \\
$\tau_{h,Na_c}$   & 1.0 & 2.0 \\
C & 0.04 & 0.08861\\
 $V_{KDR_1}$& -15 & -6.1\\
 $k_{KDR_1}$& 7.0 & 8.0\\
$n_k$ & 1 & 1 \\
   $\tau_{n,KDR_c}$    & - & 3.5 \\
 $a_{KDR}$  & 1 & - \\
 $b_{KDR}$ & 4 & - \\
$V_{KDR_2}$ & -20 & -\\
 $k_{KDR_2}$ & 7 & - \\
\Na & 2.00& 1.5\\
\IKDR & 0.5 & 0.5\\ 
  \hline
\end{tabular}
\end{center}
\end{table}
\end{center}

    \begin{figure}[!h]
\begin{center}
\centerline\leavevmode\epsfig{file=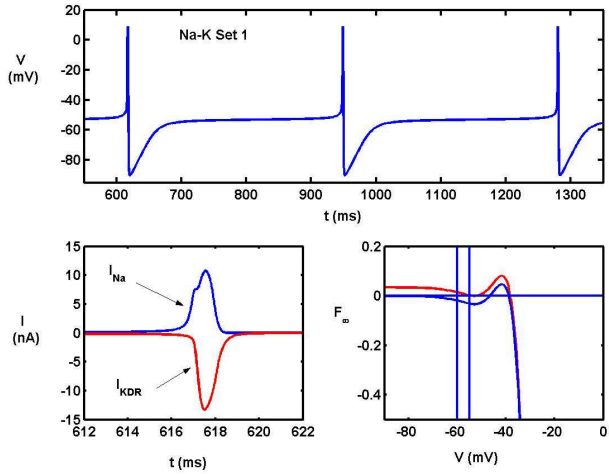,width=3.5 in}
\end{center}
\caption{Top. Repetitive spiking in the two-current model (Na-K) for the
parameter set 1 at the approximate threshold for spiking. 
Bottom left. Sodium and potassium currents during spikes.
Bottom right.  The function $F_{\infty}(V)$ defined in Equ. (16) without added current (thin blue curve) whereby spiking does not occur and with sufficient depolarizing current of 0.0342 nA (red curve) to give rise to pacemaker
activity. } 
\label{fig:wedge}
\end{figure}

Both of these  parameter sets led to repetitive spiking with the addition
of a small depolarizing current.  Typical spike trains are shown in Figures 9 and 10, 
and Table 5 contains lists of some of the details of the 
spike and spike train properties. Spiking for the second parameter 
set has a lower threshold for 
(repetitive) spiking, a longer ISI at threshold, a longer spike duration and a  larger
spike amplitude. For both parameter sets, most of these spike properties are in the ranges observed for
DRN SE neurons.
Figures  9 and 10 show well defined spikes with abruptly falling repolarization
phases to a pronounced level of hyperpolarization followed by a 
steady increase in depolarization until an apparent spike threshold is reached.
    \begin{figure}[!h]
\begin{center}
\centerline\leavevmode\epsfig{file=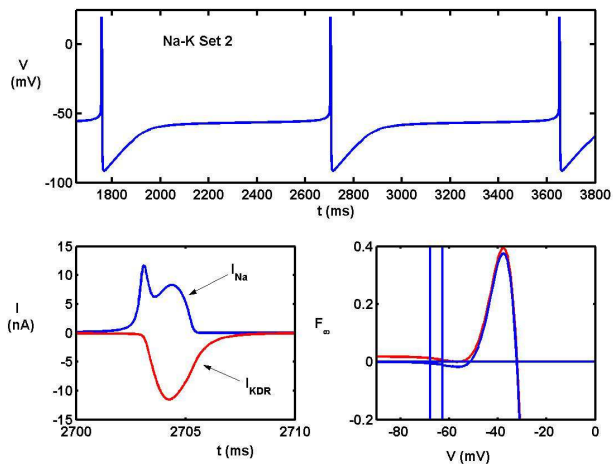,width=3.5 in}
\end{center}
\caption{Top. Repetitive spiking in the two-current model (Na-K) for the
parameter set 2 at the approximate threshold for spiking. 
Bottom left. Sodium and potassium currents during spikes.
Bottom right.  The function $F_{\infty}(V)$ defined in Equ. (16) without added current (thin blue curve) whereby spiking does not occur and with sufficient depolarizing current of 0.018 nA (red curve) to give rise to pacemaker
activity} 
\label{fig:wedge}
\end{figure}
In the lower left-hand panels of Figures 9 and 10 are shown, on an expanded time scale,  the fast sodium current $I_{Na}$ and the delayed rectifier potassium current
$I_{KDR}$ during spikes. In the lower right-hand panels are shown
plots (thin blue curves) of the function $F_{\infty}(V)$ defined as the sum of the steady state ($t \rightarrow \infty$) values of the quantities of Equ. (8) and Equ. (13),
\be F_{\infty}(V) = -[g_{Na,max} m_{Na,\infty}^3(V)h_{Na, \infty}(V)(V-V_{Na})
    +   g_{KDR,max} n_{\infty}^{n_k}(V)(V-V_{KDR})]. \ee
This behavior of this function near the resting potential has been
found to provide a heuristic indicator for spiking (Tuckwell, 2013;
Tuckwell and Penington, 2014).  It can be seen in both Figures 9 and 10
 that, for both parameter sets,  the function  $ F_{\infty}(V)$ is negative
for $V$ in an interval of considerable size around the resting potential
$V_R$ which is indicated by the vertical at -60 mV in Figure 9 and at
- 67.8 mV in Figure 10.  This means that around $V=V_R$
the derivative of $V$ with respect to time tends to be negative so that spontaneous
spiking is unlikely. The magnitude of the smallest depolarizing current required
to enable spiking is approximately the amount $-\mu$ which must
be added to make   $F_{\infty}(V) - \mu$ positive at and around
$V_R$. The resulting curves, shown in red in Figures 9 and 10,
are approximately tangential to the V-axis, being  obtained with 
$\mu=-0.0342$ for set 1 and $\mu=-0.018$ for set 2 so that these
values of $\mu$ are estimates of the threshold depolarizing current
for (pacemaker) spiking. 

DRN SE neurons often have characteristically long plateau-like
phases in the latter part of the ISI and this is apparent in Figures 9 and  10
for spikes elicited near the threshold for spiking for both sets of parameters.
The plateau for the second set is nearly three times as long as that 
for the first set. 
 In both cases (not shown) at a particular
value of $\mu$ the frequency jumps from zero to a positive value,
 being 3.0 Hz for set 1
and 1.1 Hz for set 2, so that 
this model with the chosen parameters would also  be classified as
one with Hodgkin type 2 neuron properties. 

Figure 11 gives, for both parameter sets, 
the computed spike trajectories and ISIs for levels
of excitation is not increased too much above threshold, $\mu$ being from 
1 to 1.05 times the threshold values of $\mu_c$. 
In each case the spike trajectory displays a typical pronounced 
and prolonged post-spike afterhyperpolarization followed by
a plateau-like phase before the next spike.

Although the spiking properties for the two-current (Na-K) model have
several features in common with both the experimental 
spike trains of DRN SE and LC NA neurons, with the parameter
sets in Table 4, the frequency of action potentials is  in accordance
with experimental values near threshold, but as the level of excitation
climbs to much greater values,  the frequency becomes 
somewhat high (not shown) relative to the most commonly reported  values for sustained firing in these brainstem neurons, although there are exceptions as the following examples illustrate.
In rat LC neurons, Korf et al. (1974) found in unanesthetized vivo preparations,  frequencies up to 30 Hz
and Sugiyama et al. (2012) reported in vivo frequencies over 7 hz.
In midbrain serotonergic raphe neurons, Kocsis et al. (2006) obtained in vivo rates with mean 5.4 Hz. For experiments with depolarizing
current injection, Li and Bayliss (1998) obtained initial firing rates
of around 8 Hz for caudal raphe with an injected current of 60 pA;
Li et al. (2001) reported firing of rat DRN SE cells at frequencies as
high as 35 Hz with current injection, and Ohliger-Frerking et al. (2003)
found rates as high as 8 Hz and 11 Hz with 100 pA current injection in lean and obese
Zucker rats, respectively.

     \begin{figure}[!t]
\begin{center}
\centerline\leavevmode\epsfig{file=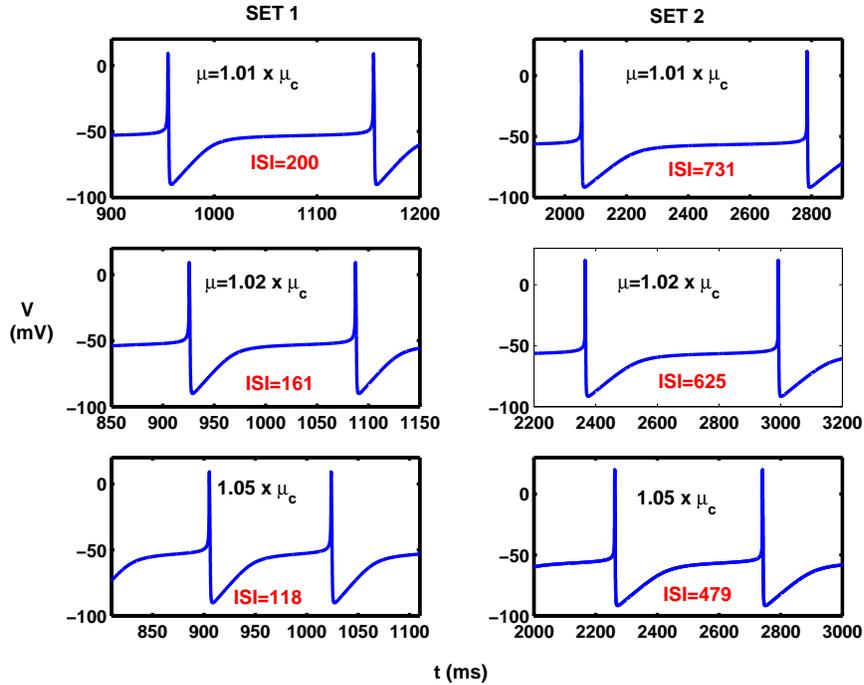,width=4.5 in}
\end{center}
\caption{Spike trajectories and ISIs obtained  in the reduced model (Na-K) for the
parameter sets of Table 4, Set 1 (left part) and Set 2 (right part) when
applied currents are increased by small amounts of 1\%. 2\% and 5\% above threshiold (top, middle, bottom curves).} 
\label{fig:wedge}
\end{figure}

\begin{center}
\begin{table}[!h]
    \caption{Spike train properties}
\smallskip
\begin{center}
\begin{tabular}{lcc}
  \hline
 Property &  Set 1 &  Basic set 2 \\
\hline 
 Spike threshold  &  $\mu=-0.0342$   &     $\mu=-0.018$ \\
ISI at threshold & 331 ms & 948 ms \\
Spike duration (-40 mV) &  1.6 ms  & 2.9 ms\\
Max V            &   +8  & +19.4 \\
Min V  &   -90.0  & -91.2 \\
  \hline
\end{tabular}
\end{center}
\end{table}
\end{center}

Hence the model is expected to be useful 
in predicting responses to random synaptic inputs in contradistinction
to a sustained depolarizing current as employed in some 
experiments,  which can lead to very large firing rates.  
However, with other parameter sets the model may not  
exhibit such high frequencies as depolarization level increases.

As Figure 1 shows, not all regular spiking in raphe and LC neurons
entails a long afterhyperpolarization and  a long plateau. Figure 11 
shows that spikes similar to those in Figures 1B, 1D and 1F are generated in the second model with increased depolarizing current.


    \begin{figure}[!ht]
\begin{center}
\centerline\leavevmode\epsfig{file=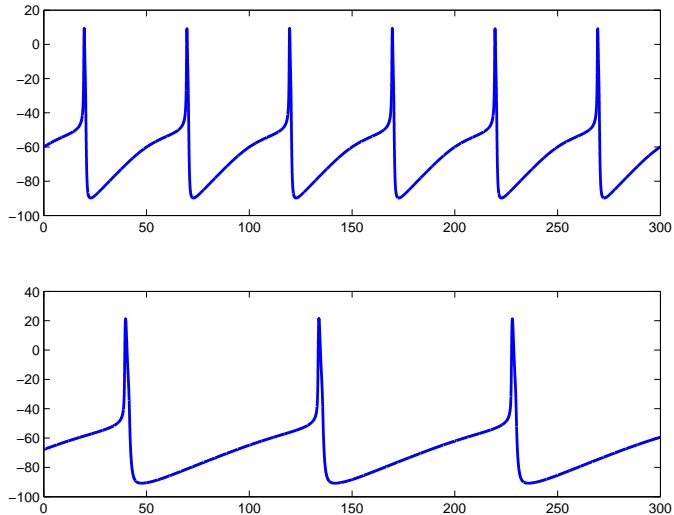,width=3.5in}
\end{center}
\caption{Spiking in the reduced model (Na-K) for the
parameter sets of Table 3 where in  both cases
$\mu=-0.05$, a value considerably above threshold and which 
results in short ISIs, characterized by relatively short plateaux and 
short-lasting AHPs.  Top part, set 1; bottom part, set 2.} 
\label{fig:wedge}
\end{figure}

\section{Discussion}

The realistic mathematical modeling of brainstem neurons, 
beyond that provided by extremely simplified models such as
the leaky integrate and fire (or Lapicque) model (Tuckwell, 1988), is needed
in order to investigate the responses of these cells to their complex array of synaptic and other input and to construct and analyze complex networks
involving these cells and those in other centers such as hippocampus,
frontal cortex and hypothalamus. 
\subsection{LC NA neurons}
Thus far there have been several mathematical models
of locus coeruleus neurons per se, which include a few ionic channels
(Putnam et al., 2014, abstract only)  or many ionic channels
including the usual sodium and potassium, high and low threshold
calcium currents, transient potassium $I_A$, persistent sodium,
leak and hyperpolarization activated cation current $I_h$ (De Carvalho et al., 2000;  Alvarez et al. 2002; 
Carter et al., 2012). Noteworthy is the omission of $I_A$ in the 
model of Alvarez et al. (2002) and its inclusion in 
De Carvalho et al. (2000) and Carter et al. (2012).  
Also, a persistent sodium current is included in De Carvalho et al. (2000) and Alvarez et al. (2002) but not in Carter et al. (2012). 

Despite such uncertainties in the mechanisms involved in
pacemaker activity in LC neurons,  some of these works have included 
synaptic input and gap-junction inputs from neighboring LC neurons.
The pioneering article of De Carvalho et al. (2000) addressed the mechanisms of morphine addiction and included 
several biochemical reactions involving cAMP, 
 $\mu$-opioid receptors,  morphine, G-protein, AC, CREB and Fos. 
Tuckwell (2015) contains a summary of previous LC modeling as well as a review of LC neuron anatomy and physiology.  
Brown et al. (2004) did employ Rose-Hindmarsh model neurons to study a network of LC neurons but 
 there have not appeared any
plausible simplified models of these cells per se. Thus the 
two-component models considered in this article provide a good starting
point for investigating, for example, the effects of synaptic input on LC firing which will be performed in future articles.  
\subsection{DRN SE neurons}
For serotonergic neurons of the dorsal raphe there has been only
one detailed model as described in the introduction (Tuckwell and
Penington, 2014). Some authors have addressed quantitatively serotonin release
and included the effects of antidepressants but 
without an explicit model for SE cell spiking (Geldof et al., 2008).
Wong-Lin et al. (2012) use a quadratic integrate and fire model for
spiking DRN SE neurons in a network of such cells along with inhibitory 
neurons. The model is not vastly different from the first
model in this article, except that the reset mechanism after spikes
is artificial. In a related work, Cano-Colino et al. (2013, 2014) have
modeled the influence of serotonin on networks of excitatory
and inhibitory cells in spatial working memory. More recently,
in a similar vein, Maia and Cano-Colino (2015) have made an interesting
study of serotonergic modulation of the strength of attractors
in orbitofrontal cortex and related this to the occurrence of
obsessive-compulsive disorder.

 \section{Concluding remarks}
Principal brainstem neurons, particularly serotonergic cells of the dorsal raphe nucleus and noradrenergic cells of the locus coeruleus are
of great importance in the functioning of many neuronal populations
throughout cortical and subcortical structures. Of note is the modulatory
role the firing of neurons in these brainstem nuclei have on neurons of the prefrontal cortex, including the 
orbitofrontal cortex, and hippocampus.  These latter structures have
been strongly implicated in various pathologies, including depression,
and obsessive-compulsive disorders. Lanfumey et al. (2008) contains
a comprehensive summary of many of the biological processes which are
influenced by serotonin including those originating
from  serotonergic neurons of the DRN. 
Modeling networks involving both serotonergic and noradrenergic afferents
requires plausible models for the spiking activity of the principal
SE and NA cells. Whereas detailed models of such activity are
now available, their application to many thousands of cells has the
disadvantage of leading to very large computation time and large memory requirements, so that  the simplified models described in the present article
may provide useful approximating  components for such complex computing tasks. 

\u

\noindent{\bf Acknowledgements} \\
This research is supported by the National Science Foundation under Agreement No. 0931642 to YZ.

%
%

\section{References}

\nh Alvarez, V.A., Chow, C.C., Van Bockstaele, E.J., Williams, J.T., 2002.
Frequency-dependent synchrony in locus ceruleus:
role of electrotonic coupling. PNAS 99, 4032-4036.

\nh Amini, B.,  Clark, J.W. Jr, Canavier, C.C., 1999. Calcium dynamics 
underlying pacemaker-like and burst
firing oscillations in midbrain dopaminergic neurons:
A computational study. J. Neurophysiol. 82, 
2249-2261.

\nh Andrade. R., Aghajanian, G.K., 1984. 
Locus coeruleus activity in vitro: intrinsic
regulation by a calcium-dependent potassium
conductance but not $\alpha_2$-adrenoceptors
J. Neurosci 4, .161-170. 

\nh Aston-Jones, G., Bloom, F.E., 1981. 
Activity of norepinephrine-containing locus coeruleus
neurons in behaving rats anticipates fluctuations in
the sleep-waking cycle. J. Neurosci. 1, 876-886. 

\nh Bayliss, D.A., Li, Y.-W., Talley, E.M., 1997a. Effects of serotonin on caudal raphe
neurons: activation of an inwardly rectifying potassium conductance. J. Neurophysiol.
77, 1349–1361. 

\nh  Belluzzi, O., Sacchi, O.,1991. A five-conductance model of the action potential
in the rat sympathetic neurone. Prog. Biophys. Molec. Biol. 55: 1-30

\nh Berridge, C.W., Waterhouse, B.D., 2003. 
The locus coeruleus–noradrenergic system: modulation of behavioral
state and state-dependent cognitive processes. Brain Res. Rev. 42, 33-84.

\nh Brown, E., Moehlis, J., Holmes, P. et al., 2004.
The influence of spike rate and stimulus duration
on noradrenergic neurons. J. Comp. Neurosci. 17, 13-29. 

\nh Cano-Colino, M., Almeida, R., Compte, A., 2013. 
Serotonergic modulation of spatial working memory: predictions from a computational network model. Frontiers in Integrative Neuroscience 7, 71. 

\nh Cano-Colino, M., Almeida, R., Gomez-Cabrero, D. et al., 2014. 
Serotonin regulates performance nonmonotonically in a spatial working
memory network. Cerebral Cortex 24, 2449-2463. 

\nh Carter, M.E., Brill, J., Bonnavion, P. et al., 2012.
Mechanism for hypocretin-mediated
sleep-to-wake transitions. PNAS 109, E2635–E2644.

\nh De Carvalho, L.A.V., De Azevedo, L.O., 2000.
A model for the cellular mechanisms
of morphine tolerance and dependence. 
Math. Comp.  Mod. 32,  933-953.

\nh De Oliveira, R.B., Howlett, M.C.H., Gravina, F.S. et al., 2010.
Pacemaker currents in mouse locus coeruleus neurons. Neurosci 170, 166-177. 

\nh De Oliveira, R.B., Gravina, F.S.,  Lim, R. et al., 2011.
Developmental changes in pacemaker currents in mouse locus
coeruleus neurons. Brain Res. 1425, 27-36. 

\nh Flower, G., Wong-Lin, K., 2014. 
Reduced computational models of serotonin
synthesis, release and reuptake.
IEEE Trans. Biomed. Eng. 61, 1054-1061.

\nh Foote, S.L., Aston-Jones, G., Bloom, F.E., 1980. 
Impulse activity of locus coeruleus neurons in awake rats and
monkeys is a function of sensory stimulation and arousal. 
Proc. Nati. Acad. Sci. USA 77, 3033-3037. 

\nh Geldof, M., Freijer,  J.I., Peletier, L.A. et al., 2008.
Mechanistic model for the acute effect of fluvoxamine on
5-HT and 5-HIAA concentrations in rat frontal cortex. 
Eur. J. Pharm. Sci. 33, 217-219.

\nh Hodgkin, A.L., 1948.  The local changes associated with repetitive action in a
non-medullated axon. J. Physiol. 107, 165-181.

 \nh  Hodgkin, A.L., Huxley, A.F., 1952.  A quantitative description of membrane
current and its application to conduction
and excitation in nerve.  J. Physiol. 117, 500-544.

\nh Jedema, H.P., Grace, A.A., 2004. 
Corticotropin-releasing hormone directly activates
noradrenergic neurons of the locus ceruleus recorded
in vitro. J. Neurosci. 24, 9703-9713.

\nh Ishikevich, E. M., 2007.
Dynamical systems in neuroscience:
the geometry of excitability and bursting. MIT Press, Cambridge, Mass.

\nh Ishimatsu, M., Williams, J.T., 1996. Synchronous activity in locus coeruleus results from dendritic
interactions in pericoerulear regions. J. Neurosci. 16, 5196-5204.

\nh Jacobs, B.L., Azmitia, E.C., 1992. Structure and function of the brain
 serotonin system.
   Physiol. Rev. 72, 165-229.

\nh  Kirby, L.G.,  Pernar, L.,  Valentino, R.J. et al., 2003. Distinguishing characteristics of serotonin and nonserotonin-
containing cells in the dorsal raphe nucleus:
electrophysiological and immunohistochemical studies. Neurosci. 116,  669-683.

\nh Kocsis B, Varga V, Dahan L, Sik A (2006)  
Serotonergic neuron diversity: Identification
of raphe neurons with discharges time-locked
to the hippocampal theta rhythm.
 PNAS 103: 1059-1064.  

\nh Korf, J., Bunney, B.S., Aghajanian, G.K., 1974.
 Noradrenergic neurons: morphine inhibition
of spontaneous activity. Eur.J. Pharmacol. 25, 165-169.

\nh Kubista, H., Boehm, S., 2006. Molecular mechanisms underlying the modulation of exocytotic
noradrenaline release via presynaptic receptors. 
Pharmacol. Therap. 112, 213-242.

\nh Lanfumey ,L., Mongeau, R., Cohen-Salmon, C., Hamon, M.,  2008.
Corticosteroid-serotonin interactions in the neurobiological mechanisms
of stress-related disorders. Neurosci. Biobehav. Rev. 32, 1174-1184.

\nh Li, Y-Q., Li, H., Kaneko, T., Mizuno, N., 2001.
Morphological features and electrophysiological properties of
serotonergic and non-serotonergic projection neurons in the dorsal
raphe nucleus
An intracellular recording and labeling study in rat brain slices. 
Brain Res. 900, 110-118.

\nh  Li, Y-W., Bayliss, D.A., 1998. Electrophysiological properties, synaptic transmission
and neuromodulation in serotonergic caudal raphe neurons. 
Clin. Exp. Pharm. Physiol.  25, 468-473.

\nh Lowry, C.A., Evans, A.K., Gasser, P.J. et al., 2008.
 Topographic organization and chemoarchitecture of
the dorsal raphe nucleus and the median raphe nucleus. In:
 Serotonin
and sleep: molecular, functional and clinical aspects,  p 25-68, 
Monti, J.M. et al., Eds.
Basel: Birkhauser Verlag AG.

\nh Luppi. P-H., Clement, O., Sapin, E. et al., 2012. 
Brainstem mechanisms of paradoxical (REM)
sleep generation. Eur. J. Physiol. 463:43-52.

\nh Maejima, T., Masseck, O.A., Mark, M.D., Herlitze, S., 2013. 
Modulation of firing and synaptic transmission of serotonergic neurons by intrinsic G protein-coupled receptors and ion channels.
Frontiers in Integrative Neuroscience 7, 40. 

\nh Maia, T.V., Cano-Colino, M., 2015.   The role of serotonin in orbitofrontal
function and obsessive-compulsive
disorder. Clinical Psychological Science 3, 460-482, and Supplemental-data. 

\nh McCormick, D.A., Huguenard, J.R., 1992. A model of the electrophysiological properties of 
thalamocortical relay neurons. J. Neurophysiol. 68, 1384-1400.

\nh Ohliger-Frerking, P.,  Horwitz, B.A., Horowitz,  J.M., 2003. 
Serotonergic dorsal raphe neurons from obese zucker
rats are hyperexcitable. Neurosci. 120, 627-634.

\nh Pan, W.J., Osmanovi\'c, S.S., Shefner, S.A., 1994.
Adenosine decreases action potential duration by modulation of
A-current in rat locus coeruleus neurons.
J. Neurosci. 14, 1114-1122. 

\nh Penington, N.J., Kelly, J.S., Fox, A.P., 1991. A study of the mechanism of Ca2+ current
inhibition produced by serotonin in rat dorsal raphe neurons. J. Neurosci. I7,
3594–3609.

\nh Putnam, R., Quintero, M., Santin, J. et al., 2014. 
Computational modeling of the effects of temperature on chemosensitive locus coeruleus neurons from bullfrogs. Faseb J. 28, Supp. 1128.3. (Abstract only).

\nh Ramirez, J-M., Koch, H., Garcia, A.J. III, et al., 2011.
The role of spiking and bursting pacemakers
in the neuronal control of breathing. J Biol. Phys. 37, 241-261.


\nh Sanchez-Padilla,J., Guzman, J.N., Ilijic, E. et al., 2014.
Mitochondrial oxidant stress in locus coeruleus is regulated by activity and nitric oxide synthase. Nat. Neurosci. 17, 832-842. 

\nh Sugiyama, D., Hur, S.W., Pickering, A:E. et al. 2012.
In vivo patch-clamp recording from locus coeruleus
neurones in the rat brainstem.
J Physiol. 590, 2225-2231.

\nh Swanson, L.W., 1976. The locus coeruleus: a cytoarchitectonic, golgi and
immunohistochemical study in the albino rat. Brain Res. 110, 39-56.

\nh Tateno,T., Harsch,A., Robinson,H.P.C., 2004. Threshold firing frequency-
current relationships of neurons in rat somatosensory cortex: Type 1
and Type 2 dynamics. J. Neurophysiol. 92, 2283-2294.

\nh Traub, R.D., Buhl, E.H., Gloveli, T., Whittington, M.A. 2003.  
Fast rhythmic bursting can be induced in layer 2/3 cortical
neurons by enhancing persistent Na$^+$ conductance
or by blocking BK channels. J. Neurophysiol. 89, 909-921.

\nh Tuckwell, H.C., 1988. Introduction to Theoretical Neurobiology. 
 Cambridge University Press, Cambridge UK. 

\nh Tuckwell, H.C., 2013. 
Biophysical properties and computational modeling of calcium spikes in
serotonergic neurons of the dorsal raphe nucleus.
BioSystems 112, 204-213. 

\nh Tuckwell, H.C., 2015. Computational modeling of spike generation in locus coeruleus noradrenergic neurons. Preprint. 

\nh Tuckwell, H.C., Penington, N.J., 2014. Computational modeling of spike
generation in serotonergic neurons of the dorsal raphe nucleus.
Prog. Neurobiol. 118, 59-101. 

\nh Vandermaelen, C.P., Aghajanian, G.K., 1983. Electrophysiological and pharmacological
characterization of serotonergic dorsal raphe neurons recorded extracellularly
and intracellularly in rat brain slices. Brain Res. 289, 109–119.

\nh Vasudeva, R.K., Waterhouse, B.D., 2014. 
Cellular profile of the dorsal raphe lateral wing sub-region:
relationship to the lateral dorsal tegmental nucleus. J. Chem. Neuroanat. 57-58,
15-23.

\nh Vertes, R.P., Crane, A.M., 1997. 
Distribution, quantification,
and morphological characteristics
of serotonin-immunoreactive cells
of the supralemniscal nucleus (b9)
and pontomesencephalic reticular
formation in the rat. J. Comp. Neurol. 378, 411-424.

\nh Williams, J.T., Egan, T.M., North, R.A., 1982. Enkephalin opens 
potassium channels on mammalian central neurons. Nature 299, 74-77. 

\nh Williams, J.T., Henderson, G., North, R.A:, 1985. 
Characterization of $\alpha_2$-adrenoceptors which
increase potassium conductance in rat locus
coeruleus neurones. Neuroscience 14, 95-101. 

\nh Wong-Lin, K-F., Joshi, A., Prasad, G., McGinnity, T.M., 2012. 
Network properties of a computational model of the dorsal raphe nucleus. Neural Netw. 32, 15-25.
%

%

\end{document}